\documentclass[twocolumn,showpacs,preprintnumbers,amsmath,amssymb,prl]{revtex4}

\usepackage{dcolumn}
\usepackage{bm}

\usepackage{latexsym}
\newcommand{\eq}{Eq}

\newcommand{\calD}{{\mathcal D}}

\newcommand{\calZ}{{\mathcal Z}}

\newcommand{\tphi}{{\widetilde\phi}}
\newcommand{\tPhi}{{\widetilde\Phi}}
\newcommand{\tJ}{{\widetilde J}}

\newcommand{\tr}{{\rm tr}}

\newcommand{\dPhi}{{\dot\Phi}}

\newcommand{\dDelta}{{\dot\Delta}}

\begin{document}

\preprint{}

\title{
Exact renormalization group approach to a nonlinear diffusion equation
}

\author{Seiichi Yoshida and Takahiro Fukui}
 \affiliation{Department of Physics, Ibaraki University, Mito
310-8512, Japan}

\date{\today}

\begin{abstract}
The exact renormalization group is applied to a nonlinear
diffusion equation with a discontinuous diffusion coefficient.
The generating functional of the solution for the initial-value 
problem of nonlinear diffusion equations
is first introduced, and next a new regularization scheme is presented. 
It is shown that the renormalization of 
an action functional in the generating functional leads to 
an anomalous diffusion exponent in full order of the perturbation
series with respect to a nonlinearity.
\end{abstract}

\maketitle

The renormalization group (RG) is a powerful tool to reveal
universal behavior of various systems, including quantum field theories
and statistical mechanics \cite{WilKog74}. 
Its basic idea lies in the coarse-graining of short-distance 
degrees of freedom, which causes redefinition of parameters governing 
the long-distance physics of the systems under investigation.
In spite of its conceptual simplicity,
there exist in the RG methods many techniques to solve with, and 
numerous applications have been made 
to equilibrium or near-equilibrium systems \cite{RGtextbook,Car96}.
In particular, exact
RG (ERG) (also called nonperturbative RG or functional RG) 
techniques and some approximations based on them 
\cite{WilKog74,WegHou73,Pol84,Wet93,Mor94}
have been attracting much renewed interest 
to reveal nonperturbative phenomena in field theories \cite{ERGrevFie}, 
statistical mechanics \cite{ERGsta}, and condensed matter physics
\cite{ZanSch,HalMet,HonSal}.

On the other hand, 
Goldenfeld {\it et al.} \cite{GMOL90,CGO} have extended the RG methods
to systems far from equilibrium. They have demonstrated that there exists a deep 
relationship between the RG and the 
intermediate asymptotics method \cite{Bar96} in the study of the nonlinear 
partial-differential equations for nonequilibrium systems.
Their idea has attracted much interest, and the RG approach to 
nonlinear differential equations has been developed \cite{RGforNLDE}.

In this paper, we apply the ERG method to a nonlinear diffusion 
equation called Barenblatt equation \cite{Bar96}. 
This equation has a discontinuous diffusion coefficient; this discontinuity
makes perturbative expansion more complicated if one proceeds to
higher order computations.
We show in this paper that Polchinski equation, a version of
the ERG equation, is very efficient 
even for such a nonlinear diffusion equation. 
It turns out that we can indeed solve 
the equation {\it for all order} in the perturbation series.
The solution leads us to the full anomalous diffusion exponent.

Let us start with the following nonlinear diffusion equation called
Barenblatt equation;
\begin{eqnarray}
&&
\partial_tu(x,t) -
D(u)\partial_x^2u(x,t)=0 
\label{BarEqu}
\end{eqnarray}
with an initial condition $u(x,t=0)=q\delta(x) $,
where $D(u)\equiv\kappa[1+g\theta(-\alpha\partial_tu)]$
denotes a nonlinear diffusion coefficient with
$\alpha$ being a positive constant which makes $\alpha\partial_t u$
dimensionless (below, we set $\alpha=1$, for simplicity). 
Here, $\theta(x)=0~(1)$ for $x<0~(x>0)$ stands for the step function. 
The dimensionless constant $g$ controls the nonlinearity of the
diffusion coefficient. 
The Barenblatt equation describes the filtration of a compressible fluid
through a compressible porous medium which can be irreversibly deformed.
Goldenfeld {\it et al.} \cite{GMOL90} have obtained  asymptotic
behavior of the solution by solving this equation via an iteration scheme
corresponding to the perturbative expansion with respect to $g$.
This perturbation gives rise to divergences:
Their basic idea is introducing a renormalization scheme 
which render the solution finite and deriving an anomalous diffusion exponent 
as an anomalous dimension in the RG language.
Though they have successfully obtained the leading correction 
of the diffusion exponent,
their method seems difficult to extend to higher order due to 
the discontinuous step-function nonlinearity.
We will present a new renormalization scheme for the initial-value problems
of nonlinear diffusion equations.

To be specific, 
we introduce a {\it generating functional} of the solution for the Barenblatt
equation and a {\it modified propagator} with a short-time cutoff 
to render the solution finite. 
To this end, notice \cite{Car96,MSR73}
that the solution of \eq. (\ref{BarEqu}) can be written as
$u(x,t)=\int\calD\phi\phi(x,t)
\delta(\partial_t\phi-D(\phi)\partial_x^2\phi)
\delta(\phi(x,0)-u(x,0))$.
This expression can be rewritten by a functional integral 
if the derivative with respect to $t$ is interpreted 
as a forward difference operator. Namely,
using the Fourier transformation for the delta function, we reach 
$u(x,t)=\langle\phi(x,t)\rangle
=\frac{1}{\calZ}\int \calD\phi\calD\tphi~\phi(x,t)e^{iS}$
with $S$ being an action functional
\begin{eqnarray}
S=\int_0^\infty \!\!\!dt\int_{-\infty}^\infty \!\!\!dx
\left[
\tphi\Delta^{-1}\phi
-g\tphi\theta(-\partial_t\phi)\kappa\partial_x^2\phi
-\tphi J
\right],
\label{ActFun}
\end{eqnarray}
where
$\Delta(x,t)\equiv e^{-x^2/(4\kappa t)}/\sqrt{4\pi\kappa t}$ 
denotes the diffusion propagator
and the generating functional $\calZ$ is defined by
$\calZ\equiv\int \calD\phi\calD\tphi~e^{iS}$, as usual. 
The field $J$ in the last term is defined by
$J(x,t)\equiv u(x,0)\delta(t)$ which controls the initial value of $u$.
In what follows, we examine the case $u(x,0)=q\delta(x)$, as mentioned 
below Eq. (\ref{BarEqu}), but it should be stressed that 
the generic initial-value problem can be treated similarly.

As discussed by Goldenfeld {\it et al.} \cite{GMOL90}, perturbative calculation 
diverges with the initial condition specified above. 
To regularize the solution, they have introduced an initial distribution 
with a finite width such as $u(x,t=0)=e^{-x^2/(2l^2)}/\sqrt{2\pi l^2}$. 
We instead introduce the short-time cutoff $\varepsilon$ for the 
propagator to formulate the ERG for the present nonlinear diffusion equation.
To be specific, we define a modified propagator as
\begin{eqnarray}
\Delta_\varepsilon(x,t)=
\theta(t-\varepsilon)\Delta(x,t).
\label{RegPro}
\end{eqnarray}
One can easily check that this propagator indeed gives a finite solution
in the perturbation theory,
applying it to the calculations 
by Goldenfeld {\it et al.} \cite{GMOL90}.
This regularization scheme can be used not only in the Barenblatt
equation but also in generic diffusion problems.

Having defined the generating functional and the modified propagator,
we next derive an ERG equation for the action functional
\eq. (\ref{ActFun}). 
Using the propagator (\ref{RegPro}) with a cutoff $\varepsilon_0$ and 
introducing a source term, we start with the generating functional,
\begin{eqnarray}
\calZ[\tJ,J]=
\int\calD\phi\calD\tphi 
e^{i\tphi\cdot\Delta^{-1}_{\varepsilon_0}\cdot\phi
-iS_{\varepsilon_0}[\tphi,\phi]-i\tJ\cdot\phi-i\tphi\cdot J} ,
\label{IniParFun}
\end{eqnarray}
where the bare action of the nonlinear term is
$S_{\varepsilon_0}[\tphi,\phi]
=g\tphi\cdot\theta(-\partial_t\phi)\kappa\partial_x^2\phi$.
At the end of the calculations, we must set $J(x,t)=q\delta(x)\delta(t)$
to obtain the solution for the initial-value problem of the present equation.
Here, the symbol $a\cdot b$ implies
$a\cdot b=\int_0^\infty dt\int_{-\infty}^\infty dx a(x,t)b(x,t)$.
Next, we introduce a new cutoff $\varepsilon~(>\varepsilon_0)$ and
divide the propagator into two parts
$\Delta_{\varepsilon_0}=\Delta_>+\Delta_<$, where
\begin{eqnarray}
&&
\Delta_>=\left[\theta(t-\varepsilon_0)-\theta(t-\varepsilon)\right]\Delta,
\nonumber\\
&&
\Delta_<=\theta(t-\varepsilon)\Delta .
\end{eqnarray}
Here, $>$ and $<$ imply the short-time and long-time modes, respectively.
Separating also $\phi$ and $\tphi$ into two fields 
$\phi=\phi_>+\phi_<$ and $\tphi=\tphi_>+\tphi_<$ enables us to
rewrite the generating functional as
\begin{eqnarray}
&&
\calZ[\tJ,J]\propto
\int\!\!\calD\phi_<\calD\tphi_<
e^{i\tphi_<\cdot\Delta_<^{-1}\cdot\phi_<}\calZ_\varepsilon
[\tJ,J;\tphi_<,\phi_<] ,
\nonumber\\
&&
\calZ_\varepsilon[\tJ,J;\tphi_<,\phi_<]=
\int
\calD\phi_>\calD\tphi_> 
e^{i\tphi_>\cdot\Delta_>^{-1}\cdot\phi_>} 
\nonumber\\
&&\mbox{}\times e^{
-iS_{\varepsilon_0}[\tphi_>+\tphi_<,\phi_>+\phi_<]
-i\tJ\cdot(\phi_>+\phi_<)-i(\tphi_>+\tphi_<)\cdot J} ,
\label{ZEpsIni}
\end{eqnarray}
up to a proportionality constant.
The field $\phi_>$, $\tphi_>$ and $\phi_<$, $\tphi_<$ can be identified 
as fields describing short-time and long-time modes, respectively.
Integrating out the short-time fields, we will next derive an effective 
action describing long-time modes.

Changing the integration variables $\tphi_>$ and $\phi_>$ into 
$\tphi_>=\tphi-\tphi_<$ and $\phi_>=\phi-\phi_<$, and 
integrating over the fields $\tphi$ and $\phi$ in \eq. (\ref{ZEpsIni}) yields
\begin{eqnarray}
\lefteqn{
\calZ_\varepsilon[\tJ,J;\tphi_<,\phi_<] }
\nonumber\\
&&=
\int\calD\tphi\calD\phi
e^{i(\tphi-\tphi_<)\cdot\Delta_>^{-1}\cdot(\phi-\phi_<)
-iS_{\varepsilon_0}[\tphi,\phi]-i\tJ\cdot\phi-i\tphi\cdot J
}
\nonumber\\
&&=
e^{-i\tJ\cdot\Delta_>\cdot J-i\tphi_<\cdot J-i\tJ\cdot\phi_<
-iS_\varepsilon[\tJ\cdot\Delta_>+\tphi_<,\Delta_>\cdot J+\phi_<]
} ,
\label{ZEpsFin}
\end{eqnarray}
where $S_\varepsilon$ is defined by 
\begin{eqnarray}
\lefteqn{
e^{-iS_\varepsilon[\tJ\cdot\Delta_>+\tphi_<,\Delta_>\cdot J+\phi_<] } }
\nonumber\\&&
\equiv
e^{-iS_{\varepsilon_0}[i\delta/\delta J+\tJ\cdot\Delta_>+\tphi_<,
i\delta/\delta\tJ+\Delta_>\cdot J+\phi_<]} .
\end{eqnarray}
This equation implies that if we
expand the exponential in r.h.s and make all the derivatives
$\delta/\delta J$ and $\delta/\delta \tJ$ 
in $S_{\varepsilon_0}$ act on $J$ and $\tJ$ in the right $S_{\varepsilon_0}$,
we reach some $S_\varepsilon$ as a functional of  
$\tPhi=\tJ\cdot\Delta_>+\tphi_<$ and $\Phi=\Delta_>\cdot J+\phi_<$.
Instead of carrying out such calculations, however, we can alternatively
determine the functional $S_\varepsilon$ by noting that 
$\calZ_\varepsilon$ obeys 
\begin{eqnarray}
\frac{d\calZ_\varepsilon}{d\varepsilon}
=
i\left(
i\frac{\delta}{\delta J}-\tphi_<
\right)
\cdot\frac{d\Delta_>^{-1}}{d\varepsilon}\cdot
\left(
i\frac{\delta}{\delta\tJ}-\phi_<
\right)\calZ_\varepsilon ,
\label{RGEquZ}
\end{eqnarray}
which follows from \eq. (\ref{ZEpsIni}). 
Substituting \eq. (\ref{ZEpsFin}) into \eq. (\ref{RGEquZ}), we obtain
the following Polchinski RG equation, 
\begin{eqnarray}
\frac{\partial S_\varepsilon}{\partial\varepsilon}
=
\frac{\delta S_\varepsilon}{\delta\Phi}
\cdot\frac{d\Delta_>}{d\varepsilon}\cdot
\frac{\delta S_\varepsilon}{\delta\tPhi}
+i\tr \frac{d\Delta_>}{d\varepsilon}\cdot
\frac{\delta^2 S_\varepsilon}{\delta\tPhi\delta\Phi} .
\label{RGEqu}
\end{eqnarray}

Next task is to determine 
the functional $S_\varepsilon$ by solving this equation.
To this end, let us first consider the $\tPhi$ dependence of the 
functional $S_\varepsilon$.
The bare $S_{\varepsilon_0}$ contains only the first order term 
in $\tPhi$, but in the process of the renormalization, 
\eq. (\ref{RGEqu}) yields the zeroth order term in $S_\varepsilon$.
To be concrete, let us denote
$S_\varepsilon[\tPhi,\Phi]=\tPhi\cdot H_\varepsilon[\Phi] +F[\Phi]$.
Substituting this into \eq.(\ref{RGEqu}), we find
\begin{eqnarray}
\frac{\partial H_\varepsilon}{\partial\varepsilon}
&=&
\frac{\delta H_\varepsilon}{\delta\Phi}
\cdot\frac{d\Delta_>}{d\varepsilon}\cdot H_\varepsilon ,
\label{RGEquH}\\
\frac{\partial F_\varepsilon}{\partial\varepsilon}
&=&
\frac{\delta F_\varepsilon}{\delta\Phi}
\cdot\frac{d\Delta_>}{d\varepsilon}\cdot H_\varepsilon 
+i\tr \frac{d\Delta_>}{d\varepsilon}\cdot
\frac{\delta H_\varepsilon}{\delta\Phi} 
\label{RGEquF}
\end{eqnarray}
with the bare functions
$H_{\varepsilon_0}[\Phi]=
g\theta(-\partial_t\Phi)\kappa\partial_x^2\Phi$ and
$F_{\varepsilon_0}[\Phi]=0$.
The RG equation for the $H_\varepsilon$-term, 
which determines the solution of the Barenblatt equation, 
is closed. Furthermore, it has no loop corrections.
Nevertheless, the initial-value problems are still nontrivial, 
since after obtaining $H_\varepsilon$, we must set $J=q\delta(x)\delta(t)$
and determine the $\varepsilon$-dependence.

To solve the functional equation (\ref{RGEquH}), we  
assume the form of $H_\varepsilon$ as 
\begin{eqnarray}
H_\varepsilon[\Phi](x,t)
\!\!&=&\!\!\int_0^\infty\!\!\! ds\int_{-\infty}^\infty\!\!\! dy
V_\varepsilon[\partial_t\Phi(x-y,t-s)]\kappa\partial_y^2\Phi(y,s)
\nonumber\\
\!\!&\equiv&\!\!
V_\varepsilon[\dPhi]\cdot \kappa\Phi''(x,t) 
\end{eqnarray}
with a certain unknown function $V_\varepsilon$,
where we have denoted $\partial_t\Phi=\dPhi$ and 
$\partial_x^2\Phi=\Phi''$ for simplicity.
Substituting this into \eq. (\ref{RGEquH}), we have
\begin{eqnarray}
\partial_\varepsilon V_\varepsilon[\dPhi]
=V_\varepsilon[\dPhi] \cdot
\partial_\varepsilon\kappa\Delta_>'' \cdot
V_\varepsilon[\dPhi] .
\label{RGEquV}
\end{eqnarray}
The bare function is given by
$V_{\varepsilon_0}[\dPhi](x-y,t-s)=-g\theta(-\dPhi(x,t))\delta(x-y)\delta(t-s)$.
This equation can be solved if $V_\varepsilon$ is expanded in power series of
$g$ such that 
$V_\varepsilon=\sum_{n=1}^\infty g^n V_\varepsilon^{(n)}$. 
Actually, we find that each term obeys
\begin{eqnarray}
\partial_\varepsilon V_\varepsilon^{(n)}[\dPhi]
=\!\!\!\!\!\!\sum_{n_1+n_2=n}V_\varepsilon^{(n_1)}[\dPhi] \cdot
\partial_\varepsilon\kappa\Delta_>'' \cdot
V_\varepsilon^{(n_2)}[\dPhi] .
\label{RGEquVn}
\end{eqnarray}
From this equation, it follows that 
$\partial_\varepsilon V^{(1)}_\varepsilon=0$ 
and hence, it turns out that $V_\varepsilon^{(1)}$ is not renormalized;
that is, $V_\varepsilon^{(1)}[\dPhi]\equiv V[\dPhi]$, where
\begin{eqnarray}
V[\dPhi](x-y,t-s)=\theta(-\dPhi(x,t))\delta(x-y)\delta(t-s) .
\label{V1}
\end{eqnarray}
This enables us to calculate the higher order solutions by
substituting \eq. (\ref{V1}) into \eq. (\ref{RGEquVn}) and by
solving it successively order by order:
\begin{eqnarray}
V_\varepsilon^{(n)}=
V\left(
\cdot\kappa\Delta_>''\cdot V\right)^{n-1} .
\label{SolV}
\end{eqnarray}
Thus, we have determined the functional $H_\varepsilon[\Phi]$ 
as infinite power series with respect to $g$.
In passing, we briefly discuss the solution for $F_\varepsilon$.
Expanding similarly the RG equation (\ref{RGEquF})  
in power series of $g$,
we find that the solution of \eq. (\ref{RGEquF}) is $F_\varepsilon=0$
because of the fact that the bare $F_{\varepsilon_0}=0$ as well as
that the second term in the r.h.s of \eq. (\ref{RGEquF}) is zero due to 
the trace with respect to the time variable. 
Therefore, we end up with the renormalized action functional,
\begin{eqnarray}
S_\varepsilon
&=&
\tPhi\cdot\sum_{n=1}^\infty g^n V[\dPhi]
\left(\cdot\kappa\Delta_>''\cdot V[\dPhi] \right)^{n-1}
\cdot\kappa\Phi'' 
\nonumber\\
&\equiv&\sum_{n=1}^\infty g^n S_\varepsilon^{(n)} 
\label{RenAct}
\end{eqnarray}
with $V$ defined by \eq. (\ref{V1}).
Thus, we have determined the renormalized action in full order in $g$.
This shows the efficiency of the present approach.

To obtain the asymptotic solution of the Barenblatt equation, 
we must set $J(x,t)=q\delta(x)\delta(t)$ to specify the initial condition
and calculate dominant parts with respect to $\varepsilon/\varepsilon_0$
in the action obtained so far.
The action decomposed into some sectors by the substitution of 
$\tPhi=\tJ\cdot\Delta_>+\tphi_<$ and $\Phi=\Delta_>\cdot J+\phi_<$.
Then, we notice that among them dominant contributions are from the  
$\tPhi(x,t)=\tphi_<(x,t)$ and $\Phi(x,t)=q\Delta(x,t)$ sectors
in the asymptotic region $\varepsilon \ll t$. 

Let us start with the first order which Goldenfeld {\it et al.} have
calculated within the perturbation in the iteration scheme.
From \eq. (\ref{RenAct}), one can obtain
\begin{eqnarray}
S_\varepsilon^{(1)}
&=&
q\int_0^\infty dt\int_{-\infty}^{\infty} dx\tphi_<(x,t)
\theta(-q\dDelta_>(x,t))\kappa\Delta_>''(x,t) 
\nonumber\\
&=&
-q\left[\frac{1}{\sqrt{2\pi e}}
\ln\frac{\varepsilon}{\varepsilon_0}\right]\tphi_<(0,0)+qS_{\rm reg}
\label{RenAct1}
\end{eqnarray}
where the regular part of $S^{(1)}$ is defined as
\begin{eqnarray}
S_{\rm reg}^{(1)}=\int_{\varepsilon_0}^\varepsilon\frac{dt}{t}
\int_{-1}^1d\omega~\delta\tphi_<(\sqrt{2\kappa t}\omega,t)~f(\omega)
\end{eqnarray}
with
$\delta\tphi(\sqrt{2\kappa t}\omega,t)\equiv
\tphi_<(\sqrt{2\kappa t}\omega,t)-\tphi_<(0,0)$
and $f(\omega)\equiv
\frac{\omega^2-1}{2}\frac{e^{-\omega^2/2}}{\sqrt{2\pi}} $.
Here, the regular term in \eq. (\ref{RenAct1}) is not involved with
the renormalization of the action since we can safely set $\varepsilon_0\to0$, 
while the first term is relevant to the
renormalization of $q$, the height of the initial distribution.
Thus it turns out that at this order $q$
is indeed renormalized, whereas others, especially $g$, is not renormalized.
This feature actually holds even in the next order, as will be checked
below. This implies that the present system is always at a fixed point 
because $g$ is not renormalized, and hence, the anomalous 
dimension which is in general a scheme-dependent quantity is a physical 
observable in the present case.
Considering these, we introduce the renormalization only to $q$ and 
define the renormalized $q$ as $q_R=qZ$. Expanding the 
renormalization constant $Z$ as
$Z=1+\sum_{n=1}g^nZ^{(n)}$ with 
$Z^{(n)}=-\gamma^{(n)}\ln(\varepsilon/\varepsilon_0)$,
it turns out that the first order of $Z$ reads
\begin{eqnarray}
\gamma^{(1)}=\frac{1}{\sqrt{2\pi e}} ,
\end{eqnarray}
from \eq. (\ref{RenAct1}). This indeed reproduces the result of 
Goldenfeld {\it et al.} \cite{GMOL90}.

The renormalization of $q$ introduced above is indeed enough to
render the solution of the Barenblatt equation finite also in higher order. 
To verify this,
let us next calculate the second order renormalized action.
Due to similar arguments to the first order, the action (\ref{RenAct})
yields
\begin{eqnarray}
S^{(2)}_\varepsilon
&=&
q\int_{2\varepsilon_0}^{ \varepsilon_0+\varepsilon }\!\!\!dt
\int_{-\sqrt{2\kappa t}}^{\sqrt{2\kappa t}} \!\!\!dx
\int_{\varepsilon_0}^{t-\varepsilon_0}\!\!\!dt_1
\int_{-\sqrt{2\kappa t_1}}^{\sqrt{2\kappa t_1}} \!\!\!dx_1
\nonumber\\ &&\times 
\tphi_<(x,t)
\kappa\Delta''(x-x_1,t-t_1)\kappa\Delta''(x_1,t_1)
\nonumber\\&&
+\mbox{reg. } ,
\end{eqnarray}
where reg. stands for regular parts of the renormalized action.
Similar but a bit lengthy calculation leads to 
\begin{eqnarray}
S_\varepsilon^{(2)}
&=&
q\frac{Z^{(1)2}}{2}
\tphi(0,0)
+qZ^{(1)}S_{\rm reg}^{(1)}
+qZ^{(2)}\tphi(0,0)
\nonumber\\&&  
+\mbox{reg.} 
\end{eqnarray}
where $Z^{(2)}=-\gamma^{(2)}\ln(\varepsilon/\varepsilon_0)$ with
\begin{eqnarray}
\lefteqn{       
\gamma^{(2)}=-
\int_0^1\frac{d\tau_1}{\tau_1}
\int_{-1}^1d\omega_1 f(\omega_1)
}\nonumber\\ &\times&  
\left[\frac{1}{\sqrt{2\pi e}}
-\frac{1-\sqrt{\tau_1}\omega_1}{1-\tau_1}
\frac{e^{-(1-\sqrt{\tau_1}\omega_1)^2/\left[2(1-\tau_1)\right]}}
{\sqrt{2\pi(1-\tau_1)}}
\right] .
\label{AnoDim2}
\end{eqnarray}
This result indicates that the second order action correctly includes
the contributions from the first order renormalized action. 
Namely, the renormalized action with the source term $q\tphi(0,0)$
satisfies 
$gS^{(1)}_\varepsilon+g^2S^{(2)}_\varepsilon+q\tphi(0,0)
=e^{Z}qS_{\rm reg}+e^{Z}q\tphi(0,0)$ 
up to the second order of $g$.
Therefore, we expect in general that
$
q_R=q\left(\varepsilon_0/\varepsilon\right)^{\gamma}
$
where
\begin{eqnarray}
\gamma=\sum_{n=1}g^n\gamma^{(n)}.
\label{AnoDim}
\end{eqnarray}
The constant $\gamma$ thus obtained indeed gives the anomalous dimension
of the solution for the present diffusion equation.

To see this, notice that since $\tJ$ is not renormalized, we have
$u(x,t;q,\varepsilon_0)=u(x,t;q_R,\varepsilon)$,
which tells that the solution is independent of $\varepsilon$.
Hence, the renormalized solution should satisfy the following RG equation;
\begin{eqnarray}
\left(
\varepsilon\frac{\partial}{\partial\varepsilon}-
\gamma q_R\frac{\partial}{\partial q_R}
\right)u(x,t;q_R,\varepsilon)=0,
\label{RenQ}
\end{eqnarray}
where $\gamma$ is defined by \eq. (\ref{AnoDim}), or alternatively by
$\gamma=-\partial \ln q_R/\partial \ln\varepsilon$.
On the other hand, the dimensional analysis requires
$u(x,t;q,\varepsilon_0)=
q/\sqrt{\kappa t}\Phi
\left(x/\sqrt{\kappa t}, \varepsilon_0/t\right)$.
Therefore, combining these observations, we can assume
$u(x,t;q_R,\varepsilon)=q_R/\sqrt{\kappa t}\Phi
\left(x/\sqrt{\kappa t}, \varepsilon/t\right)$.
Substituting this into \eq. (\ref{RenQ}), it turns out that the relation
$\Phi\left(x/\sqrt{\kappa t}, \varepsilon/t\right)=
\left(\varepsilon/t\right)^\gamma
\tilde\Phi\left(x/\sqrt{\kappa t}\right)$ 
holds. Hence, the asymptotic behavior of 
the solution for the Barenblatt equation is indeed given by
$u\sim 1/t^{1/2+\gamma}$.

So far we have derived the anomalous diffusion exponent up to 
second order with respect to $g$. The exponent of first order is 
the same as that obtained by Goldenfeld {\it et al.}, whereas
the exponent of second order (\ref{AnoDim2}) is estimated as
$\gamma^{(2)}\sim-0.063546$ by numerical integration.
Though  this value is different from the result in Ref. 
\cite{GMOL90}, 
it coincides with that obtained later by Cole and 
Wagner given by a different integral formula derived via different method
\cite{ColWag96}. Since the anomalous diffusion exponent should be 
scheme-independent in the present case,
we believe that our result as well as Cole and Wagner's is correct.

The efficiency of our method lies in the fact that one can compute the 
higher order exponent in a similar way above. 
To present the exponent of $n$th order,
we define a function,
\begin{eqnarray}
g(\omega_1,\omega_2,\tau_2)=
\frac{e^{-(\omega_1-\sqrt{\tau_2}\omega_2)^2/[2(1-\tau_2)]}}
{\sqrt{2\pi(1-\tau_2)}}
-\frac{e^{-\omega_2^2/2}}{\sqrt{2\pi}} .
\end{eqnarray}
Then, \eq. (\ref{RenAct}) yields
\begin{eqnarray}
\gamma^{(n)}&=&-\int_0^1 \prod_{j=1}^{n-1}\frac{d\tau_j}{\tau_j}
\int_{-1}^1\prod_{j=1}^{n-1}d\omega_jf(\omega_{n-1}) 
\nonumber\\ &\times& 
\left[\frac{1}{\sqrt{2\pi e}}
-\frac{1-\sqrt{\tau_1}\omega_1}{1-\tau_1}
\frac{e^{-(1-\sqrt{\tau_1}\omega_1)^2/\left[2(1-\tau_1)\right]}}
{\sqrt{2\pi(1-\tau_1)}}
\right] 
\nonumber\\ &&\mbox{ }\times
\prod_{j=1}^{n-2}
\frac{1}{2}\frac{d^2 g(\omega_{j},\omega_{j+1},\tau_{j+1})}{d\omega_j^2}.
\end{eqnarray}
We expect that this anomalous exponent is {\it exact}, since it has been 
derived from one formula (\ref{RenAct}), among which the first and second 
order coincides with those calculated by different methods.
For references, 
we numerically estimate the exponent of third order,
$\gamma^{(3)}=-0.00314$.
The exponent obtained so far seems a good convergent series.

In summary,
we have applied the ERG techniques to the initial-value problem of the
Barenblatt equation, one of typical nonlinear diffusion equations.
We have derived the anomalous diffusion exponent in full 
order with respect to the parameter controlling the nonlinearity.
This implies that the ERG approach is efficient to systems far from 
equilibrium described by nonlinear partial-differential
equations as well as to 
field theories and statistical mechanics.

The present approach would be useful to other types of initial-value problem
for nonlinear diffusion equations. In particular, application to 
the critical dynamic of nonlinear traveling wave is of great interest.
One of well known examples is the KPP equation \cite{KPP} which shows an
interesting universal behavior in the selection of the front velocity.  
This problem has been addressed by 
Paquette {\it et al.} \cite{PCGO94}, but more detailed analysis is needed if 
one wants to understand, e.g., the universal logarithmic corrections
to the velocity in the pulled-front, from the RG point of view.
Application of the ERG to such problems would be quite interesting.

One of the author (TF) would like to thank T. Ohta and T. Kunihiro 
for valuable discussions. This work was supported in part
by  Grant-in-Aid for Scientific Research from JSPS.



\end{document}